\documentclass[pra, twocolumn, superscriptaddress, amsmath, amssymb]{revtex4-2}
\usepackage[utf8]{inputenc}
\usepackage{amsmath}
\usepackage{physics}
\usepackage{amsfonts}
\usepackage{amssymb}
\usepackage{graphicx}
\usepackage{color}
\usepackage{txfonts}
\usepackage{float}
\usepackage[titletoc]{appendix}
\usepackage[colorlinks={true}]{hyperref}
\hypersetup{
    citecolor={blue},
    filecolor={blue},
    linkcolor={blue},
    urlcolor={blue}
}
\begin{document}
    \title{Liouvillian spectral control for fast charging of quantum batteries}

    \author{Hang Zhou}
    \thanks{These authors contributed equally to this work.}
    \affiliation{Hunan Key Laboratory of Nanophotonics and Devices, Hunan Key Laboratory of Super-Microstructure and Ultrafast Process, and School of Physics, Central South University, Changsha 410083, China}

    \author{Jia-Wei Huang}
    \thanks{These authors contributed equally to this work.}
    \affiliation{Hunan Key Laboratory of Nanophotonics and Devices, Hunan Key Laboratory of Super-Microstructure and Ultrafast Process, and School of Physics, Central South University, Changsha 410083, China}

    \author{Chuan-Cun Shu}
    \email{cc.shu@csu.edu.cn}
    \affiliation{Hunan Key Laboratory of Nanophotonics and Devices, Hunan Key Laboratory of Super-Microstructure and Ultrafast Process, and School of Physics, Central South University, Changsha 410083, China}

    \begin{abstract}
        Quantum batteries, which use quantum systems to store and deliver energy,
        are promising for next-generation energy storage. However, optimizing charging
        strategies and understanding the interplay between dissipation and quantum
        coherence remain open challenges. Here, we investigate steady-state charging
        in an open quantum battery and demonstrate that the charging timescale depends
        on the spectral gap of the Liouvillian operator governing dissipative dynamics.
        As a minimal example, we examine a three-level quantum battery realized in
        a single trapped ${}^{40}\mathrm{Ca}^{+}$ ion, where energy from an
        engineered thermal photon reservoir is coherently transferred to a long-lived
        metastable storage state. We find that long-term dynamics are confined to
        a low-dimensional manifold of slow Liouvillian modes, with their
        spectral structure determining the relaxation rate to the charged
        steady state. By adjusting experimentally accessible parameters,
        such as reservoir occupation and coherent coupling strength, the non-Hermitian
        Liouvillian spectrum can approach an exceptional point. This increases
        the spectral gap and accelerates the approach to steady state. As a result,
        this mechanism significantly enhances asymptotic charging power without
        relying on many-body collectivity or steady coherence. Our findings offer
        fundamental insights into open quantum thermodynamics and provide a path
        to efficient energy storage and fast-charging solutions in emerging quantum
        technologies.
    \end{abstract}

    \maketitle

    \section{Introduction}
    \label{Introduction} Quantum batteries (QBs) use quantum resources such as
    coherence and entanglement, offering the potential to exceed classical performance
    limits   \cite{PhysRevE.87.042123,Binder_2015,PhysRevLett.118.150601,PhysRevLett.120.117702,RevModPhys.96.031001}.
    Recent progress in trapped ions, superconducting circuits, and solid-state
    devices has enabled experimental demonstrations and benchmarking of QB
    charging protocols   \cite{Quach3160,Hu_2022,sp5l-c6m8,Hymas2026,g45c-ssfx}. However,
    because practical QBs are open quantum systems subject to dissipation and
    decoherence, it remains unclear how to fully harness quantum effects to improve
    practical energy efficiency   \cite{PhysRevA.102.052223,PhysRevA.109.022226,PhysRevApplied.14.024092,PhysRevE.102.062133,PhysRevResearch.7.013151,PhysRevLett.134.180401,43n6-rnj3}.
    Two main strategies have emerged: one leverages coherence and entanglement to
    temporarily boost charging power or reduce energy costs   \cite{PhysRevA.106.032212,PhysRevA.108.052213,PhysRevA.106.062609,PhysRevLett.111.240401,PhysRevLett.129.130602,s6dl-zgkx,PhysRevB.104.245418},
    while the other aims for superlinear improvements through collective many-body
    dynamics
      \cite{4klp-kw27,PhysRevA.105.022628,PhysRevA.109.032201,PhysRevA.111.022222,PhysRevApplied.19.064069,PhysRevB.105.115405,PhysRevB.109.235432,PhysRevE.99.052106,PhysRevE.111.044118,PhysRevLett.133.243602,PhysRevResearch.4.043150}.\\
    \indent
    For open QBs, the main challenge is understanding and controlling charging as
    the system interacts with its environment. Unlike closed QBs, where transient
    quantum effects may dominate, practical QBs must address noise and loss, which
    affect the long-term relaxation rate toward steady states. Recent studies
    show that dissipation and decoherence can stabilize energy and improve
    robustness, shifting attention to open dynamics that govern long-term charging   \cite{PhysRevLett.122.210601,PhysRevLett.132.090401,PhysRevE.104.044116,PhysRevResearch.2.013095,Shastri2025,Chang_2021}. In this approach, the spectral gap of the Liouvillian superoperator is critical, as it determines
    the slowest relaxation timescale and directly impacts charging speed   \cite{Spohn1977,PhysRevLett.116.240404,PhysRevA.89.022118,PhysRevA.100.062131,PhysRevA.101.062112,PRXQuantum.2.040346,PhysRevA.106.012207,PhysRevLett.108.070604,PhysRevLett.122.110601,PhysRevLett.128.110402,PhysRevLett.130.110402,PhysRevLett.132.243602,Bu2024,Zhang2022,v66j-6fv4,PhysRevResearch.5.043036,PhysRevA.98.042118,Ashida02072020}.
    However, systematic Liouvillian spectral engineering for dissipative QB charging remains largely unexplored.

    In this work, we present a Liouvillian spectral-control strategy for steady-state
    charging in open QBs. Performance is defined by the steady
    energy and relaxation timescale, which together determine average charging
    power. As an example, we use a single trapped $^{40}\mathrm{Ca}^{+}$ ion, selected
    for its established control techniques, clear level structure, and long-lived
    metastable states suitable for energy storage. By systematically adjusting
    reservoir occupation and coherent coupling strength, we control the Liouvillian
    spectrum. This targeted approach reorganizes relaxation pathways and
    significantly increases the spectral gap, accelerating the approach to steady
    state. Our method enables efficient charging in dissipative environments
    without relying on complex many-body effects or fragile quantum resources, supporting
    optimized energy storage in practical open quantum systems.\\
    \indent
    The paper is organized as follows: Section~\ref{model} introduces the three-level
    QB model and its Lindblad master equation. Section~\ref{EPs} analyzes the slow-mode
    manifold, Liouvillian spectrum, and gap structure. Section~\ref{results}
    connects spectral properties to steady-state charging and demonstrates
    relaxation dynamics acceleration through simulations. Section~\ref{Experimental_scope}
    discusses an implementation with a trapped $^{40}\mathrm{Ca}^{+}$ ion.
    Finally, Section~\ref{Conclusion} presents the conclusions.\\
    \indent
    \section{Theoretical methods}
    \subsection{Dissipative quantum battery charging model }
    \label{model}
\begin{figure}[h]
        \centering
        \includegraphics[width=1\linewidth]{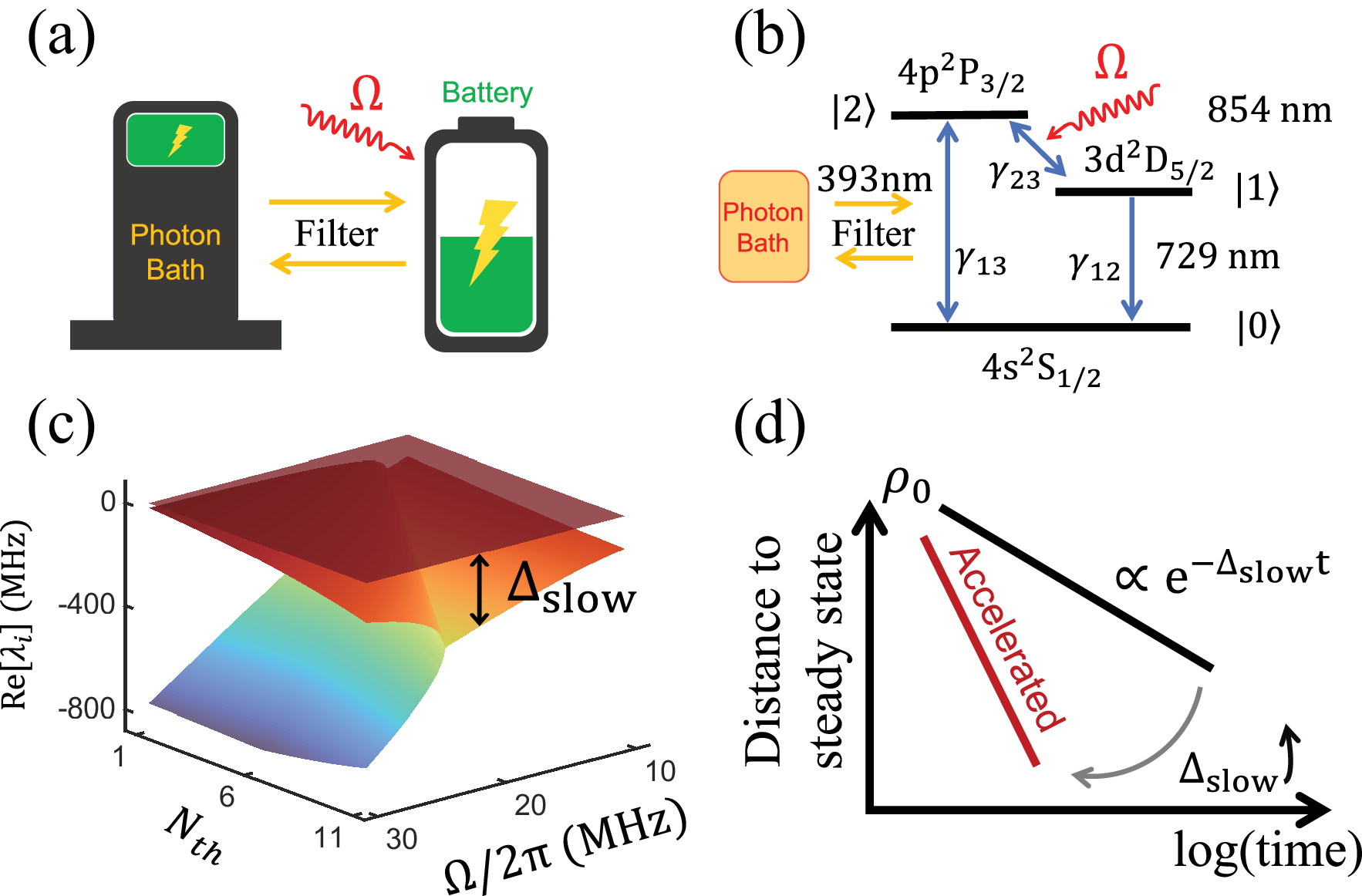}
        \caption{Dissipative quantum battery (QB) charging. (a) Schematic of a QB
        connected to a thermal photon reservoir via a frequency-selective filter
        and driven by a coherent control field ($\Omega$). (b) Level structure
        of $^{40}\mathrm{Ca}^{+}$, showing the ground state $|0\rangle = 4s^{2}S_{1/2}$,
        metastable storage state $|1\rangle = 3d^{2}D_{5/2}$, and short-lived excited
        state $|2\rangle = 4p^{2}P_{3/2}$. Controlled charging is achieved
        through dissipative transitions ($\gamma_{21}$ spontaneous decay at 854 nm
        and $\gamma_{20}$ spontaneous decay at 393 nm) and coherent coupling ($\Omega$
        at 854 nm). (c) Liouvillian spectrum, indicating steady (zero), slow
        (near the origin), and fast-decaying modes. (d) Relaxation dynamics showing
        the Hilbert–Schmidt norm to the steady state, determined by the slowest
        Liouvillian modes.}
        \label{fig1}
    \end{figure}

We theoretically investigate a minimal three-level open quantum
    battery, as depicted in Fig.~\ref{fig1}. The  model is representative
    of an experimental setup based on a single trapped ${}^{40}\mathrm{Ca}^{+}$ ion.
    The working medium consists of the ground state
    $\lvert 0\rangle=4s^{2}S_{1/2}$, a metastable state $\lvert 1\rangle=3d^{2}D_{5/2}$
    with a lifetime of several seconds, and a short-lived excited state
    $\lvert 2\rangle=4p^{2}P_{3/2}$ with a nanosecond lifetime. Energy is supplied
    via the optical transition between $\lvert 0\rangle$ and $\lvert 2\rangle$,
    which is coupled to an engineered thermal photon reservoir with mean
    occupation number $N_{\mathrm{th}}$. A resonant continuous-wave control field
    drives the transition from $\lvert 2\rangle$ to $\lvert 1\rangle$  at Rabi frequency
    $\Omega$, coherently transferring population to the long-lived storage level
    $\lvert 1\rangle$.\\
    \indent

    The total Hamiltonian of the system is given by ($\hbar=1$)
    \begin{equation}
        \hat{H}(t) = \hat{H}_{\mathrm{B}}+ \hat{H}_{\mathrm{S}}+ \hat{H}_{\mathrm{SB}}
        + \hat{H}_{\mathrm{C}}(t),
    \end{equation}
    where $\hat{H}_{\mathrm{B}}= \sum_{i=0}^{2}\omega_{i}|i\rangle \langle i|$ describes
    the three-level QB, and
    $\hat{H}_{\mathrm{S}}= \sum_{k}\nu_{k}a_{k}^{\dagger}a_{k}$ represents the
    bosonic reservoir. The reservoir couples to the battery only through the $|0\rangle
    \leftrightarrow |2\rangle$ transition. The system-bath interaction is given
    by
    $\hat{H}_{\mathrm{SB}}= \sum_{k}g_{k}(|0\rangle\langle 2| + |2\rangle\langle
    0|)(a_{k}+ a_{k}^{\dagger})$, which accounts for both thermally stimulated excitation
    and spontaneous emission. Coherent control is applied via a classical
    driving field resonantly coupling the $|1\rangle$ and $|2\rangle$ levels, described
    by $\hat{H}_{\mathrm{C}}(t) = (\mu_{12}\mathcal{E}(t))(|1\rangle\langle 2|
    + |2\rangle\langle 1|)$, where
    $\mathcal{E}(t) = \mathcal{E}_{0}\cos(\omega_{L}t)$. Moving to the interaction
    picture with respect to $\hat{H}_{\mathrm{B}}+ \hat{H}_{\mathrm{S}}$ and applying
    the rotating-wave approximation, the coherent dynamics are governed by the
    effective Hamiltonian
    $\hat{H}_{I}= \Omega (|1\rangle\langle 2| + \mathrm{H.c.}) + \delta |2\rangle
    \langle 2|$, where the detuning is $\delta = \omega_{21}- \omega_{L}$ and the
    Rabi frequency is $\Omega = \mu_{12}\mathcal{E}_{0}/2$.

    The reduced density operator $\rho(t)$ obeys the Lindblad master equation~  \cite{10.1063/1.522979,Lindblad1976}
    (see Appendix~\ref{supp1})
    \begin{equation}
        \frac{d\rho}{dt}=-i[\hat{H}_{I},\rho]+\sum_{j}\mathcal{D}
        _{j}[\rho],
    \end{equation}
    with dissipators
    \begin{align}
        \mathcal{D}_{20}[\rho] & =\gamma_{20}(N_{\mathrm{th}}+1)\,\mathcal{D}\!\left[\lvert 0\rangle\langle 2\rvert\right]\rho+\gamma_{20}N_{\mathrm{th}}\,\mathcal{D}\!\left[\lvert 2\rangle\langle 0\rvert\right]\rho,\nonumber \\[2pt]
        \mathcal{D}_{21}[\rho] & =\gamma_{21}\,\mathcal{D}\!\left[\lvert 1\rangle\langle 2\rvert\right]\rho,\label{eq:dissip}                                                                                                     \\[2pt]
        \mathcal{D}_{10}[\rho] & =\gamma_{10}\,\mathcal{D}\!\left[\lvert 0\rangle\langle 1\rvert\right]\rho,\nonumber
    \end{align}
    where $\gamma_{20}$, $\gamma_{21}$, $\gamma_{10}$ denote
    spontaneous decay rates for $\lvert 2\rangle\rightarrow\lvert 0\rangle$,
    $\lvert 2\rangle\rightarrow\lvert 1\rangle$, and $\lvert 1\rangle\rightarrow\lvert
    0\rangle$, respectively. The Lindblad superoperator is $\mathcal{D}[A]\rho=A\rho
    A^{\dagger}-\tfrac{1}{2}\{A^{\dagger}A,\rho\}$. Projecting onto the basis
    $\{\lvert 2\rangle,\lvert 1\rangle,\lvert 0\rangle\}$ yields
    \begin{align}
        \dot{\rho}_{22} & =-i\Omega(\rho_{12}-\rho_{21})+\gamma_{20}\!\left[N_{\mathrm{th}}\rho_{00}-(N_{\mathrm{th}}+1)\rho_{22}\right]-\gamma_{21}\rho_{22},\nonumber       \\
        \dot{\rho}_{11} & =i\Omega(\rho_{12}-\rho_{21})+\gamma_{21}\rho_{22}-\gamma_{10}\rho_{11},\nonumber                                                                   \\
        \dot{\rho}_{00} & =\gamma_{20}\!\left[(N_{\mathrm{th}}+1)\rho_{22}-N_{\mathrm{th}}\rho_{00}\right]+\gamma_{10}\rho_{11},\label{eq:rho0}                               \\
        \dot{\rho}_{12} & =i\delta\rho_{12}+i\Omega(\rho_{11}-\rho_{22})-\tfrac{1}{2}\!\left[\gamma_{10}+\gamma_{21}+\gamma_{20}(N_{\mathrm{th}}+1)\right]\rho_{12},\nonumber
    \end{align}
    subject to $\rho_{22}+\rho_{11}+\rho_{00}=1$. These equations describe coherent
    transfer within the  manifold $\{\lvert 1\rangle,\lvert 2\rangle\}$ and reservoir-assisted
    exchange between $\lvert 0\rangle$ and $\lvert 2\rangle$. Efficient
    channeling of excitations from $\lvert 2\rangle$ to the metastable state
    $\lvert 1\rangle$ enables population accumulation in the storage level, realizing
    QB charging.

    To characterize steady-state charging, we define the relaxation time
    $\tau_{\mathrm{s}}$ as the earliest time satisfying
    $\|\rho(\tau_{\mathrm{s}})-\rho_{\mathrm{ss}}\|<\varepsilon$, where $\rho_{\mathrm{ss}}$ 
    is the stationary solution satisfying $\dot{\rho}=0$, $\varepsilon$ is a convergence
    threshold, and $\|A\|=\sqrt{\mathrm{Tr}(AA^{\dagger})}$ denotes the Hilbert--Schmidt
    norm. For an initially discharged battery $\rho_{0}=|0\rangle\langle0|$, the steady stored energy and average charging power are defined as
    \begin{equation}
        E_{\mathrm{s}}=\mathrm{Tr}\!\left(\rho_{\mathrm{ss}}\hat{H}_{\mathrm{B}}\right
        ),\qquad P_{\mathrm{s}}=\frac{E_{\mathrm{s}}}{\tau_{\mathrm{s}}},
    \end{equation}
    where the QB ground-state energy is set to zero. Thus $P_{\mathrm{s}}$
    represents the mean rate of energy accumulation during the approach to
    stationarity, distinct from instantaneous power.

    \subsection{ Liouvillian spectral control}
    \label{EPs}
   By vectorizing the density operator as $|\rho\rangle$ in Liouville space, the steady-state behavior of the driven dissipative battery can be 
fully determined by 
\begin{equation}
\frac{d|\rho\rangle}{dt}=\mathcal{L}|\rho\rangle,\quad \mathcal{L}|\rho_{ss}\rangle=0
\end{equation}
where Liouvillian  superoperator $\mathcal{L}$ is non-Hermitian in operator space and has complex eigenvalues, which determine the relaxation spectrum of the open QB.
In the ordered basis of the three-level battery, the full Liouvillian
superoperator $\mathcal{L}$ is given by   \cite{PhysRevA.100.062131,PhysRevA.101.062112}
\begin{equation}
\begin{aligned}
\mathcal{L}=&-i\left(\hat{H}_I\otimes \mathcal{I}-\mathcal{I}\otimes \hat{H}_I^T\right)\\ 
&+\sum{\mu}\left[J\mu\otimes  J_\mu^*-\frac{1}{2}\left(J_\mu^\dagger J_\mu\otimes \mathcal{I}+\mathcal{I}\otimes (J_\mu^\dagger J_\mu)^T\right)\right],
\end{aligned}
\end{equation}
where $\mathcal{I}$ denotes the identity operator in the battery Hilbert space. The jump operators $J_\mu$ represent reservoir-induced excitation and decay channels. In this model, $J_\mu$ includes $\sqrt{\gamma_{20}(N_{\rm th}+1)}|0\rangle\langle2|$, $\sqrt{\gamma_{20}N_{\rm th}}|2\rangle\langle0|$, $\sqrt{\gamma_{21}}|1\rangle\langle2|$, and $\sqrt{\gamma_{10}}|0\rangle\langle1|$.

Introducing right and left eigenoperators $|R_{\alpha}\rangle$
    and $\langle L_{\alpha}|$ satisfying
    \begin{equation}
        \mathcal{L}|R_{\alpha}\rangle=\lambda_{\alpha}|R_{\alpha}\rangle, \qquad
        \langle L_{\alpha}|\mathcal{L}=\lambda_{\alpha}\langle L_{\alpha}|,
    \end{equation}
    and normalized through $\langle L_{\alpha}|R_{\beta}\rangle=\delta_{\alpha\beta}$,
    the density operator admits the spectral expansion
    \begin{equation}
        |\rho(t)\rangle = |\rho_{\mathrm{ss}}\rangle + \sum_{\alpha\ge1}c_{\alpha}
        e^{\lambda_\alpha t}|R_{\alpha}\rangle,
    \end{equation}
    where $\lambda_{0}=0$ corresponds to the unique steady state and
    $c_{\alpha}=\langle L_{\alpha}|\rho(0)\rangle$. For primitive Markovian dynamics,
    all nonzero eigenvalues satisfy $\mathrm{Re}[\lambda_{\alpha}]<0$, implying that
    the long-time deviation from stationarity decays exponentially as
    \begin{equation}
        \|\rho(t)-\rho_{\mathrm{ss}}\| \sim e^{-\Delta t}, \qquad \Delta = -\max_{\alpha\neq0}
        \mathrm{Re}[\lambda_{\alpha}].
    \end{equation}
    The Liouvillian gap $\Delta$ therefore sets the intrinsic relaxation time scale
    $\tau_{\Delta}=1/\Delta$ that limits steady-state charging.

    For the three-level QB in this work, relaxation time scales are separated in the experimentally relevant limit.
\begin{equation}
\gamma_{10}\ll {\gamma_{20},\gamma_{21},\Omega}.
\end{equation}
In this regime, the metastable storage state decays much more slowly than both optical decay and driven population-transfer processes. As a result, long-time relaxation arises only from the sector involving populations and coherences directly coupled by the control field. To clarify this structure, we vectorize the density matrix and reorder the Liouville-space basis as $|\rho\rangle =(\rho_{22},\rho_{12},\rho_{21},\rho_{11},\rho_{00},\rho_{20},\rho_{10},\rho_{02},\rho_{01})^{\mathrm T}$. With this ordering, the Liouvillian separates into
\begin{equation}
\mathcal{L}\simeq\mathcal{L}_{5}\oplus\mathcal{L}_{2}^{(L)}\oplus\mathcal{L}_{2}^{R} .
\end{equation}
The five-dimensional block $\mathcal{L}_{5}$ acts on $(\rho_{22},\rho_{12},\rho_{21},\rho_{11},\rho_{00})^{\mathrm T}$, which includes the populations of the three levels  and the coherence between  the driven manifold ${|1\rangle,|2\rangle}$. This block is the only sector that contributes to stored energy and contains the slow modes governing the approach to the charged steady state, as shown in Appendix~\ref{supp2}. The two $2\times2$ blocks, $\mathcal{L}_{2}^{(L)}$ and $\mathcal{L}_{2}^{R}$, act on $(\rho_{20},\rho_{10})^{\mathrm T}$ and $(\rho_{02},\rho_{01})^{\mathrm T}$, respectively, and describe coherences involving the ground state. Their eigenvalues have real parts set by the optical reservoir coupling, of order $\gamma_{20}(2N_{\mathrm{th}}+1)$. These modes decay rapidly and do not affect asymptotic relaxation.

Therefore, in the limit $\gamma_{10}\ll {\gamma_{20},\gamma_{21},\Omega}$, the long-time charging dynamics is governed by the reduced slow-sector equation
\begin{equation}
\frac{d}{dt}|\rho_{5}\rangle
=
\mathcal{L}_{5}|\rho_{5}\rangle .
\end{equation} The slow-sector gap relevant for the slow-time approach to the steady state
    is
    \begin{equation} \label{Slow}
        \Delta_{\mathrm{slow}}=-\max_{\beta\neq 0}\mathrm{Re}[\lambda_{\beta}(\mathcal{L}
        _{5})].
    \end{equation}
    Under the above hierarchy, $\Delta_{\mathrm{slow}}$ also determines the
    overall Liouvillian gap, as all remaining modes decay on parametrically shorter
    time scales.  This reduced slow-mode structure and the corresponding
    asymptotic relaxation toward the steady state are illustrated
    schematically in Fig.~\ref{fig1}(c) and Fig.~\ref{fig1}(d), respectively.\\
    \indent
The reduced generator $\mathcal{L}_{5}$ is intrinsically non-Hermitian and displays spectral degeneracies typical of open quantum systems. As the experimentally tunable parameters $N_{\mathrm{th}}$ and $\Omega$ are varied, the leading eigenvalues of $\mathcal{L}_{5}$ reorganize qualitatively. In
    particular, a complex-conjugate pair of slow eigenvalues emerges,
    \begin{equation}
        \lambda_{\pm}=-\gamma_{\mathrm{eff}}\pm i\omega ,
    \end{equation}
    where $\gamma_{\mathrm{eff}}>0$ is the effective
    decay rate and $\omega$ is the oscillation frequency
    of the slow manifold. As parameters are tuned, this pair approaches the real
    axis and coalesces at a Liouvillian exceptional point (EP), beyond which it splits
    into two distinct real eigenvalues $-\gamma_{1}$ and $-\gamma_{2}$ ($\gamma_{1,2}
    >0$)  \cite{Heiss_2012}. This transition marks a crossover from underdamped oscillatory
    relaxation to overdamped monotonic convergence toward the steady state.

  The Liouvillian gap can be locally enhanced near the EP due to the
spectral reorganization of the dominant slow modes. As shown in Appendix~\ref{supp2}, the eigenvalues of
$\mathcal{L}_{5}$ satisfy a cubic characteristic equation whose vanishing
discriminant identifies the coalescence of two slow eigenvalues. Together
with the rank condition
$\dim\ker(M-\lambda_{\mathrm{EP}}I)=1$, this coalescence corresponds to a
defective slow-sector degeneracy, namely a second-order Liouvillian EP. Near this degeneracy the dominant pair assumes the universal square-root
    form
    \begin{equation}
        \lambda_{\pm}= -\gamma_{\mathrm{eff}}\pm \sqrt{\kappa - \kappa_{\mathrm{EP}}}
        ,
    \end{equation}
    where the control parameter $\kappa=\kappa(N_{\mathrm{th}},\Omega)$
    characterizes the effective coupling within the slow sector and
    $\kappa_{\mathrm{EP}}$ denotes its critical value. For $\kappa < \kappa_{\mathrm{EP}}$
    the square root is imaginary and the dynamics is underdamped; for
    $\kappa > \kappa_{\mathrm{EP}}$ it is real and both eigenvalues become
    purely dissipative. According to Eq.~(\ref{Slow}), the relevant gap is set by the dominant
nonzero eigenvalue of \(\mathcal{L}_5\). In the parameter regime considered
here, this dominant mode belongs to the coalescing eigenvalue pair, while
the remaining nonzero slow eigenvalue remains spectrally separated. Evaluation of the cubic roots in Appendix~\ref{supp2}
shows that, when the EP is approached from the underdamped side, the real
part of the dominant pair moves away from zero, increasing
$\Delta_{\mathrm{slow}}$. On the overdamped side, the real splitting
generates one branch with a real part closer to zero, thereby reducing
$\Delta_{\mathrm{slow}}$. Thus, in the present model and along the
considered tuning path, the EP marks an effective critical-damping condition
for the slow manifold, where the asymptotic relaxation toward the steady
state is fastest.

    The slow manifold has a clear physical origin in the battery architecture.
    Energy is injected via reservoir-induced transitions between $\ket{0}$ and
    $\ket{2}$, while it is stored in the long-lived metastable state $\ket{1}$.
    This separation yields distinct time scales: fast reservoir-driven processes
    and slow storage dynamics. Eliminating the fast coherences leaves a reduced
    set of dynamical variables controlling the asymptotic approach to the
    charged steady state. Tuning parameters reshapes the non-Hermitian spectrum
    of this reduced generator, providing a mechanism for accelerating steady-state
    charging through spectral engineering rather than many-body collectivity or
    steady coherence.

    \section{Results and discussion}
    \label{results}We conduct numerical simulations for the three-level $^{40}\mathrm{Ca}^{+}$ ion with $\omega_{0}=0$, $\omega_{1}=1.70$ eV, and $\omega_{2}=3.15$ eV.
These values correspond to transition frequencies $\omega_{20}=3.15$ eV, $\omega_{10}=1.70$ eV, and $\omega_{21}=1.45$ eV. Spontaneous emission rates are $\gamma
    _{20}=140$ MHz, $\gamma_{21}=9$ MHz, and $\gamma_{10}\approx1.3\times10^{-6}$
    MHz   \cite{NIST_ASD_2024,PhysRevA.62.032503,Gerritsma2008,PhysRevLett.111.023004,PhysRevLett.115.143003}.
    We solve the Lindblad master equation in the time domain and analyze the
    Liouvillian spectrum in the full nine-dimensional operator space. \\
    \indent
    \subsection{Slow-manifold control of asymptotic charging}
    \label{Results_A}
    \begin{figure}[h]
        \centering
        \includegraphics[width=1\linewidth]{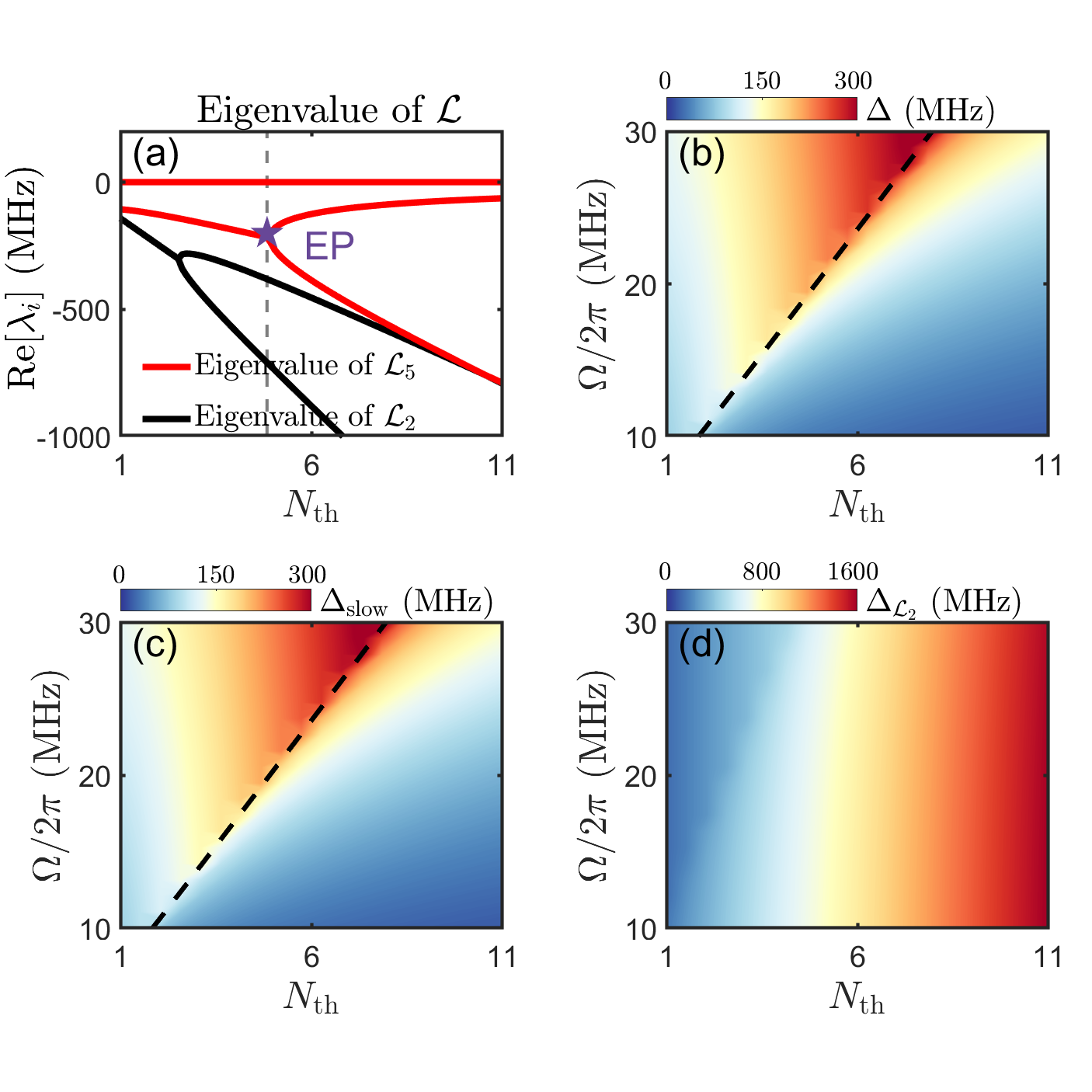}
        \caption{\label{fig2} Liouvillian spectral properties as functions of
        the thermal photon occupation ($N_{\mathrm{th}}$) and the coherent drive
        strength ($\Omega$). (a) Real parts of the dominant nonzero eigenvalues of
        the full Liouvillian $\mathcal{L}$ as a function of $N_{\mathrm{th}}$ at
        fixed $\Omega$ $/2\pi=20~\mathrm{MHz}$. Panel (b) shows the Liouvillian
        gap $\Delta$ in the $(N_{\mathrm{th}},\Omega)$ plane. Panel (c) displays
        the slow-sector gap $\Delta_{\mathrm{slow}}$ in the $(N_{\mathrm{th}},\Omega
        )$ plane. Panel (d) presents the gap $\Delta_{\mathcal{L}_2}$ associated
        with the branch $\mathcal{L}_{2}$ in the same parameter space. Dashed lines
        in panels (b) and (c) indicate the location of the EP, as extracted from
        panel (a). The detuning is $\delta=0$ in all panels.}
    \end{figure}
    To determine the spectral origin of the asymptotic charging dynamics, we
    examine how the Liouvillian structure changes with thermal occupation
    $N_{\mathrm{th}}$ and coherent drive strength $\Omega$, as shown in Fig.~\ref{fig2}.
    Figure~\ref{fig2}(a) shows a clear separation in the Liouvillian spectrum
    between a slow branch $\mathcal{L}_{5}$ and a faster branch
    $\mathcal{L}_{2}$. As $N_{\mathrm{th}}$ increases, two eigenvalues in the
    slow branch converge, resulting in an EP that signals a reorganization of the
    dominant slow manifold. Figures~\ref{fig2}(b) and \ref{fig2}(c) show that
    the full Liouvillian gap $\Delta$ and the slow-sector gap
    $\Delta_{\mathrm{slow}}$ have the same structure across the
    $(N_{\mathrm{th}},\Omega)$ plane. In contrast, Fig.~\ref{fig2}(d) shows that
    the gap $\Delta_{\mathcal{L}_2}$ for the faster branch remains much larger
    and changes smoothly across parameter space. These spectral features reveal a
    clear separation of timescales. Fast modes decay quickly and do not affect
    late-time evolution, while the slow manifold determines the asymptotic relaxation
    rate. The ridge of maximal $\Delta =\Delta_{\mathrm{slow}}$ identifies the
    parameter regime where long-time charging dynamics are fastest.\\
    \indent
    \begin{figure}[h]
        \centering
        \includegraphics[width=0.95\linewidth]{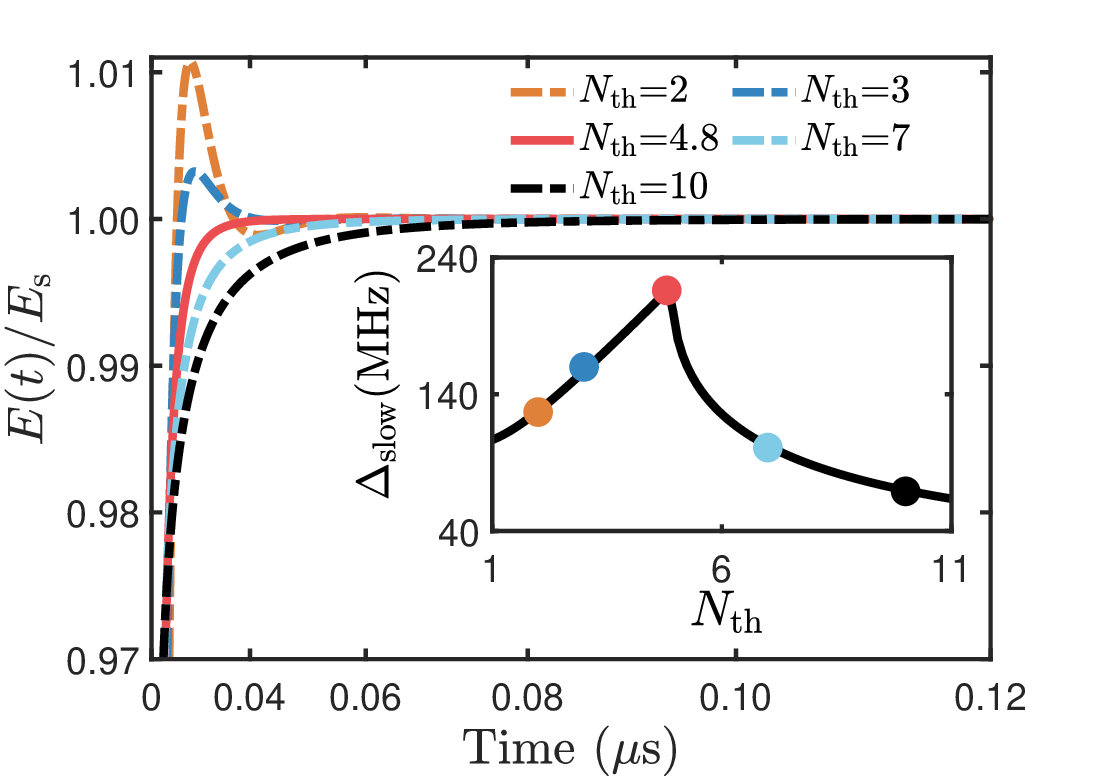}
        \vspace{-6pt}
        \caption{\label{fig3}Time evolution of the stored energy, normalized to
        its steady-state value $E(t)/E_{s}$, for various thermal photon occupations
        $N_{\mathrm{th}}$. The inset displays the slow-sector gap
        $\Delta_{\mathrm{slow}}$ as a function of $N_{\mathrm{th}}$, with colored
        markers indicating the parameter values shown in the main panel. The
        parameters for the simulations are $\delta/2\pi = 0$ and
        $\Omega/2\pi = 20~\mathrm{MHz}$.}
        \vspace{-6pt}
    \end{figure}
    Figure~\ref{fig3} illustrates the time evolution of the normalized stored
    energy $E(t)/E_{s}$ for different values of $N_{\mathrm{th}}$. The convergence
    rate to the steady value is governed by the thermal occupation number:
    it rises with increasing $N_{\mathrm{th}}$, reaches a maximum near $N_{\mathrm{th}}
    \approx 4.8$, and then declines as $N_{\mathrm{th}}$ increases further. At low
    thermal occupation, the charging curve slightly overshoots $E_{s}$, whereas at
    higher $N_{\mathrm{th}}$, the approach to the steady state is monotonic.
    The inset reveals that this transition aligns with the parameter range where
    $\Delta_{\mathrm{slow}}$ attains its maximum. This change in charging
    dynamics signifies a reorganization of the slow-sector spectrum. The overshoot
    observed at low $N_{\mathrm{th}}$ reflects underdamped relaxation, driven by
    a pair of dominant slow eigenvalues with complex conjugate components. In this
    regime, coherent transfer into the storage level temporarily surpasses the
    ultimate dissipative redistribution. As $N_{\mathrm{th}}$ increases, these
    eigenvalues become real, leading to a crossover to overdamped relaxation and
    a more direct approach to the steady state.

    \begin{figure}[h]
        \centering
        \includegraphics[width=1\linewidth]{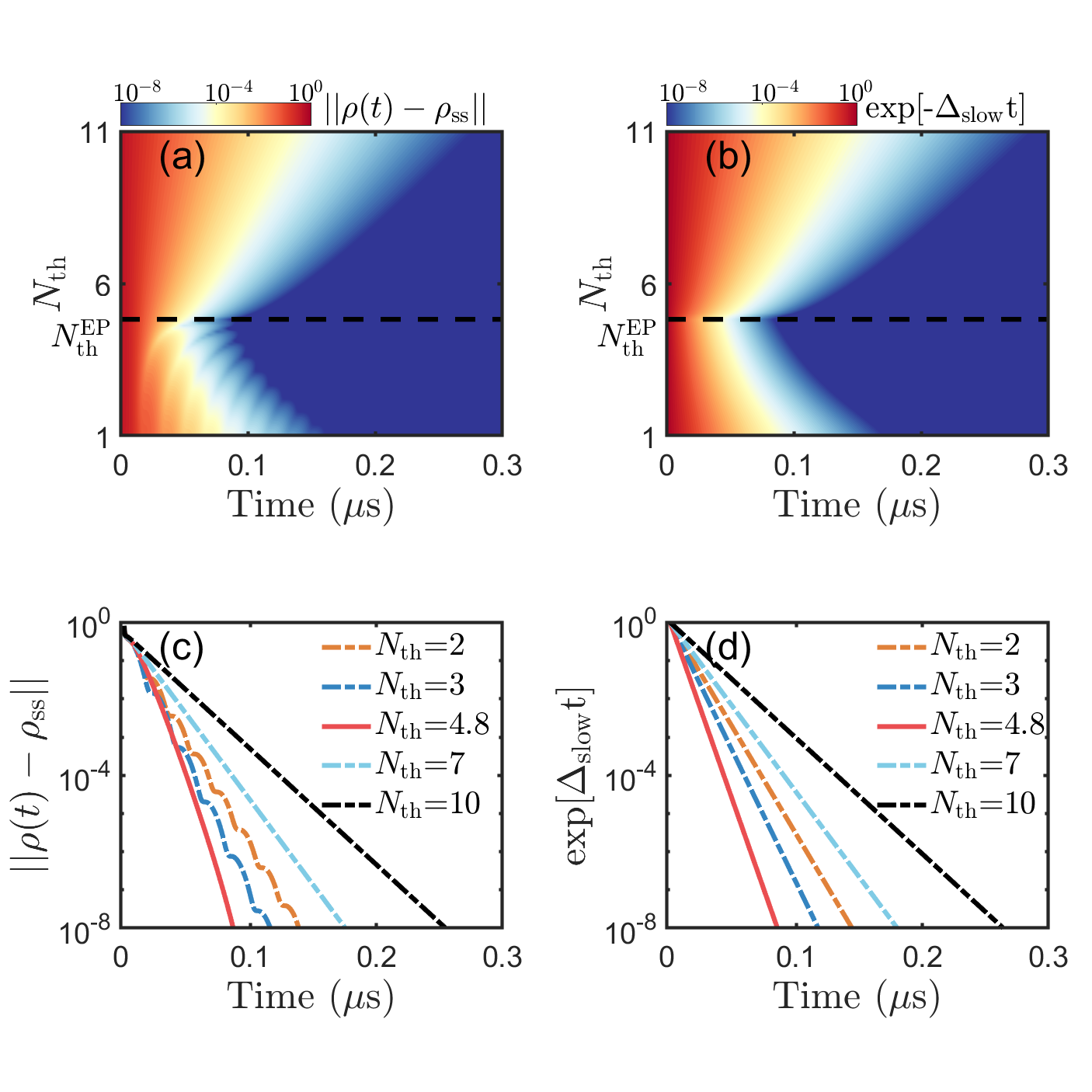}
        \caption{\label{fig4} Hilbert–Schmidt norm to the steady state as a
        function of time and thermal photon occupation. (a) Hilbert–Schmidt norm
        $\|\rho(t)-\rho_{\mathrm{ss}}\|$ versus time for various thermal occupations
        $N_{\mathrm{th}}$. (b) Corresponding exponential reference curves
        $\exp[-\Delta_{\mathrm{slow}}t]$. (c) Hilbert–Schmidt norm $\|\rho(t)-\rho
        _{\mathrm{ss}}\|$ for selected $N_{\mathrm{th}}$. (d) Corresponding
        exponential curves $\exp[-\Delta_{\mathrm{slow}}t]$ for the same
        parameter values. The horizontal dashed line in (a) and (b) indicates $N_{\mathrm{th}}
        ^{\mathrm{EP}}$. The other parameters are the same as in Fig.~\ref{fig3}.}
    \end{figure}
    Figure~\ref{fig4} analyzes the relaxation dynamics by tracking the Hilbert–Schmidt
    norm to the steady state, $\|\rho(t)-\rho_{\mathrm{ss}}\|$, and comparing
    it to the reference exponential $\exp[-\Delta_{\mathrm{slow}}t]$. In Fig.~\ref{fig4}(a),
    decay for $N_{\mathrm{th}}$ below $N_{\mathrm{th}}^{\mathrm{EP}}$ exhibits pronounced
    oscillations, while above $N_{\mathrm{th}}^{\mathrm{EP}}$, the decay is smooth
    and monotonic. Figure~\ref{fig4}(b) shows the corresponding exponential reference
    curves, and Figs.~\ref{fig4}(c) and \ref{fig4}(d) provide representative line
    cuts comparing the exact Hilbert–Schmidt norm with the exponential trends.
    Although transient dynamics differ across EP, the long-time decay consistently
    follows the exponential form determined by $\Delta_{\mathrm{slow}}$. The
    results in Figs.~\ref{fig2}--\ref{fig4} demonstrate that the slow-sector gap
    governs the asymptotic approach to the steady charged state, with early-time
    deviations arising from subleading modes and mode interference. The spectral
    structure of the slow Liouvillian manifold determines the asymptotic
    charging dynamics.

    \subsection{Gap-controlled charging performance}
    \label{Results_B}
    \begin{figure}[h]
        \centering
        \includegraphics[width=1\linewidth]{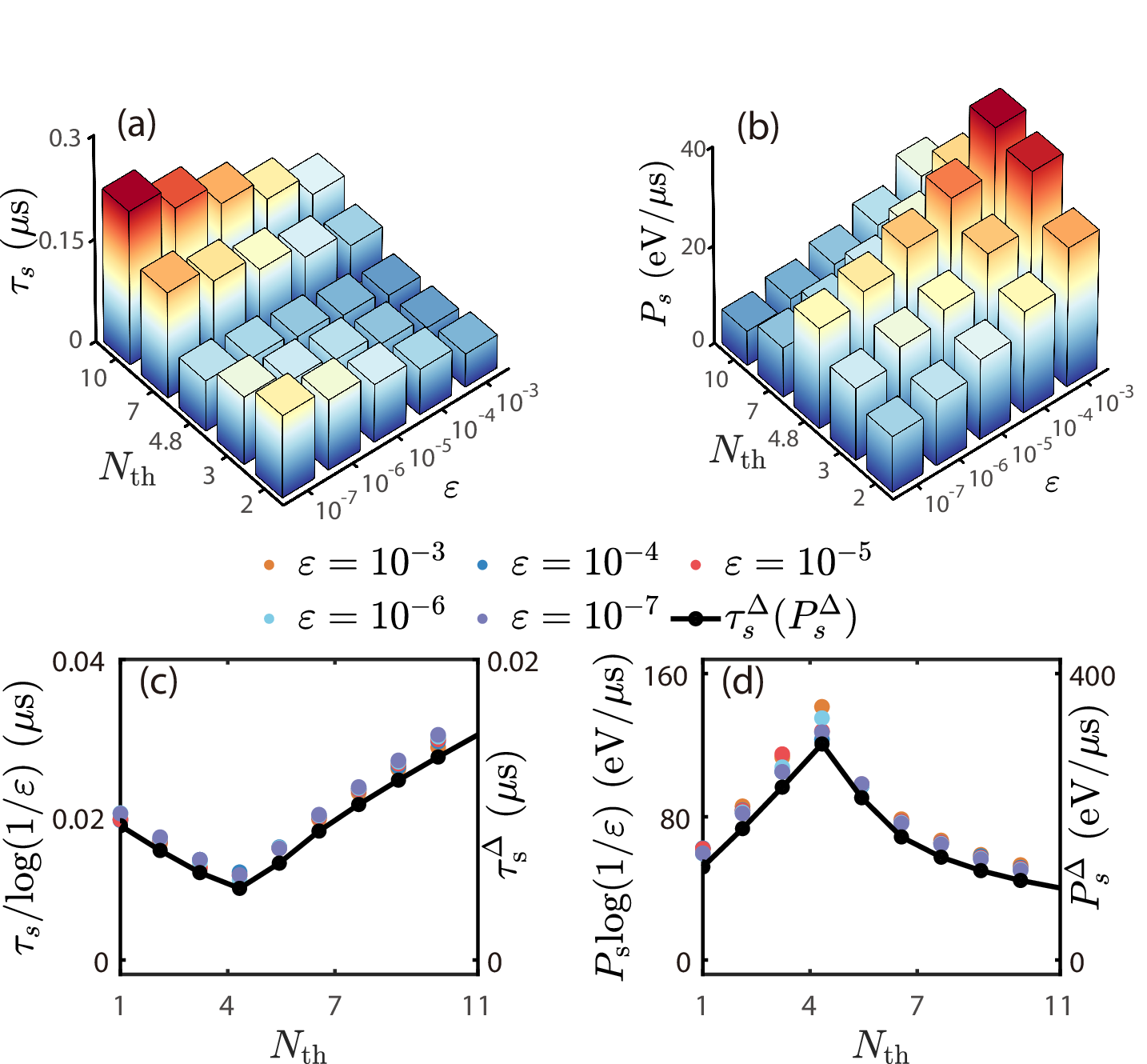}
        \caption{\label{fig5} Dependence of steady-state charging time and
        charging power on the threshold parameter $\varepsilon$. (a) steady-state
        charging time $\tau_{s}$ as a function of thermal photon occupation $N_{\mathrm{th}}$
        and threshold $\varepsilon$. (b) steady-state charging power $P_{s}$
        as a function of $N_{\mathrm{th}}$ and $\varepsilon$. (c) Rescaled charging
        time $\tau_{s}/\log(1/\varepsilon)$ versus $N_{\mathrm{th}}$ for representative
        values of $\varepsilon$ (left axis), with the gap-based reference
        $\tau_{s}^{\Delta}$ shown on the right axis. (d) Rescaled power $P_{s}\log
        (1/\varepsilon)$ versus $N_{\mathrm{th}}$ for the same $\varepsilon$ values
        (left axis), with the gap-based reference $P_{s}^{\Delta}$ shown on the
        right axis. The other parameters are the same as in Fig.~\ref{fig3}.}
    \end{figure}
    To determine whether the observed charging enhancement depends on the
    definition of saturation, Fig.~\ref{fig5} shows how the extracted charging
    time $\tau_{s}$ and steady-state charging power $P_{s}$ vary with the
    convergence threshold $\varepsilon$. As shown in Figs.~\ref{fig5}(a) and \ref{fig5}(b),
    changing $\varepsilon$ affects the absolute values of $\tau_{s}$ and $P_{s}$,
    but the nonmonotonic relationship with $N_{\mathrm{th}}$ remains consistent
    across several orders of magnitude in $\varepsilon$. In Figs.~\ref{fig5}(c) and
    \ref{fig5}(d), the rescaled quantities $\tau_{s}/\log(1/\varepsilon)$ and $P_{s}
    \log(1/\varepsilon)$ closely follow the gap-based reference values $\tau_{s}^{\Delta}$
    and $P_{s}^{\Delta}$. The optimal regime's position is largely unchanged as
    $\varepsilon$ varies. These results show that the convergence threshold mainly
    introduces a logarithmic scaling to the extracted charging time and power,
    without affecting their dependence on $N_{\mathrm{th}}$. Therefore, the
    observed enhancement in charging performance is an intrinsic feature of the
    Liouvillian relaxation dynamics, not an artifact of the operational definition
    of saturation.\\
    \indent
    \begin{figure}[h]
        \centering
        \includegraphics[width=1\linewidth]{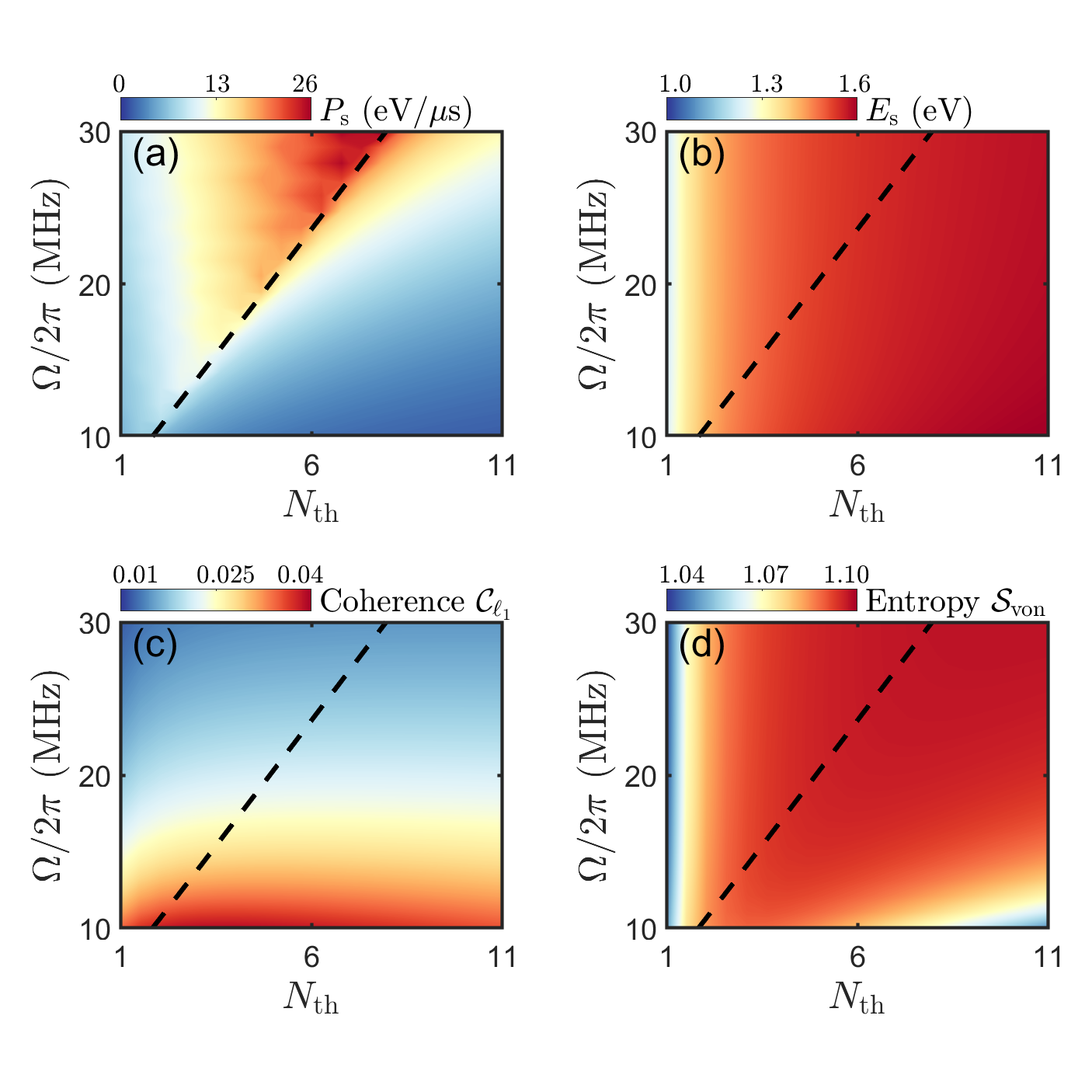}
        \caption{\label{fig6}Steady-state properties of the QB in the $(N_{\mathrm{th}}
        ,\Omega)$ parameter plane. (a) Steady-state charging power
        $P_{\mathrm{s}}$. (b) Stored energy $E_{\mathrm{s}}=\mathrm{Tr}(\rho_{\mathrm{ss}}
        \hat H_{B})$. (c) $\ell_{1}$ coherence $C_{\ell_1}=\sum_{i\neq j}|\rho_{ij}|$ of
        the steady state in the energy eigenbasis. (d) Von Neumann entropy $S
        _{\mathrm{von}}=-\mathrm{Tr}(\rho_{\mathrm{ss}}\ln\rho_{\mathrm{ss}})$.
        The dashed lines indicate the location of the exceptional point. The parameters
        for the simulations are $\delta/2\pi = 0$ and $\varepsilon = 10^{-8}$.}
    \end{figure}
    This behavior results from the spectral decomposition that governs the long-term
    evolution of the system
    \begin{equation}
        |\rho(t)\rangle=e^{\mathcal{L}t}|\rho_{0}\rangle=|\rho_{\rm ss}\rangle+\sum_{\alpha\ge1}c_{\alpha}
        e^{\lambda_\alpha t}|R_{\alpha}\rangle,
    \end{equation}
    where $\rho_{0}$ is the initial density operator and $\mathrm{Re}[\lambda_{1}]=-\Delta$ represents the slowest nonzero decay
    rate. For typical initial states with nonzero overlap with this mode, the long-term
    deviation from the steady state is primarily determined by the envelope $e
    ^{-\Delta t}$, leading to
    \begin{equation}
        \tau_{s}(\varepsilon)\simeq \frac{1}{\Delta}\ln\!\left(\frac{1}{\varepsilon}
        \right).
    \end{equation}
    Near a second-order EP, the merging of slow modes introduces only minor corrections
    due to the Jordan-block structure, while the dominant $1/\Delta$ scaling
    remains. \\
    \indent
    Figure~\ref{fig6} compares several steady-state QB properties across the $(N_{\mathrm{th}}
    ,\Omega)$ parameter space. In Fig.~\ref{fig6}(a), charging power $P_{s}$ forms
    a pronounced ridge at intermediate thermal occupation and strong driving,
    closely following the exceptional-point trajectory. In contrast, Fig.~\ref{fig6}(b)
    shows that steady stored energy $E_{s}$ increases smoothly with
    $N_{\mathrm{th}}$ and depends only weakly on $\Omega$. Figure~\ref{fig6}(c)
    demonstrates that steady $\ell_{1}$ coherence
    $C_{\ell_1}=\sum_{i\neq j}|\rho_{ij}|$ peaks at weak driving and remains low
    in the high-power region   \cite{PhysRevLett.113.140401}. Fig.~\ref{fig6}(d) indicates
    that von Neumann entropy is elevated near the exceptional-point line, but
    similarly high entropy also appears in regions with much lower charging
    power.\\
    \indent
    These comparisons show that the high-power ridge does not correspond to
    increased stored energy, steady coherence, or mixedness of the
    steady state. The observed power enhancement is not due to higher final energy
    or stronger steady-state quantum coherence. Instead, its alignment with the exceptional-point
    trajectory suggests the enhancement results from a spectral reorganization of
    slow-relaxation modes, specifically a dynamical increase in the Liouvillian
    gap, rather than from static properties of the steady state.
    \subsection{Sensitivity analysis of gap-controlled charging}
    \label{Results_C}
    \begin{figure}[h]
        \centering
        \includegraphics[width=1\linewidth]{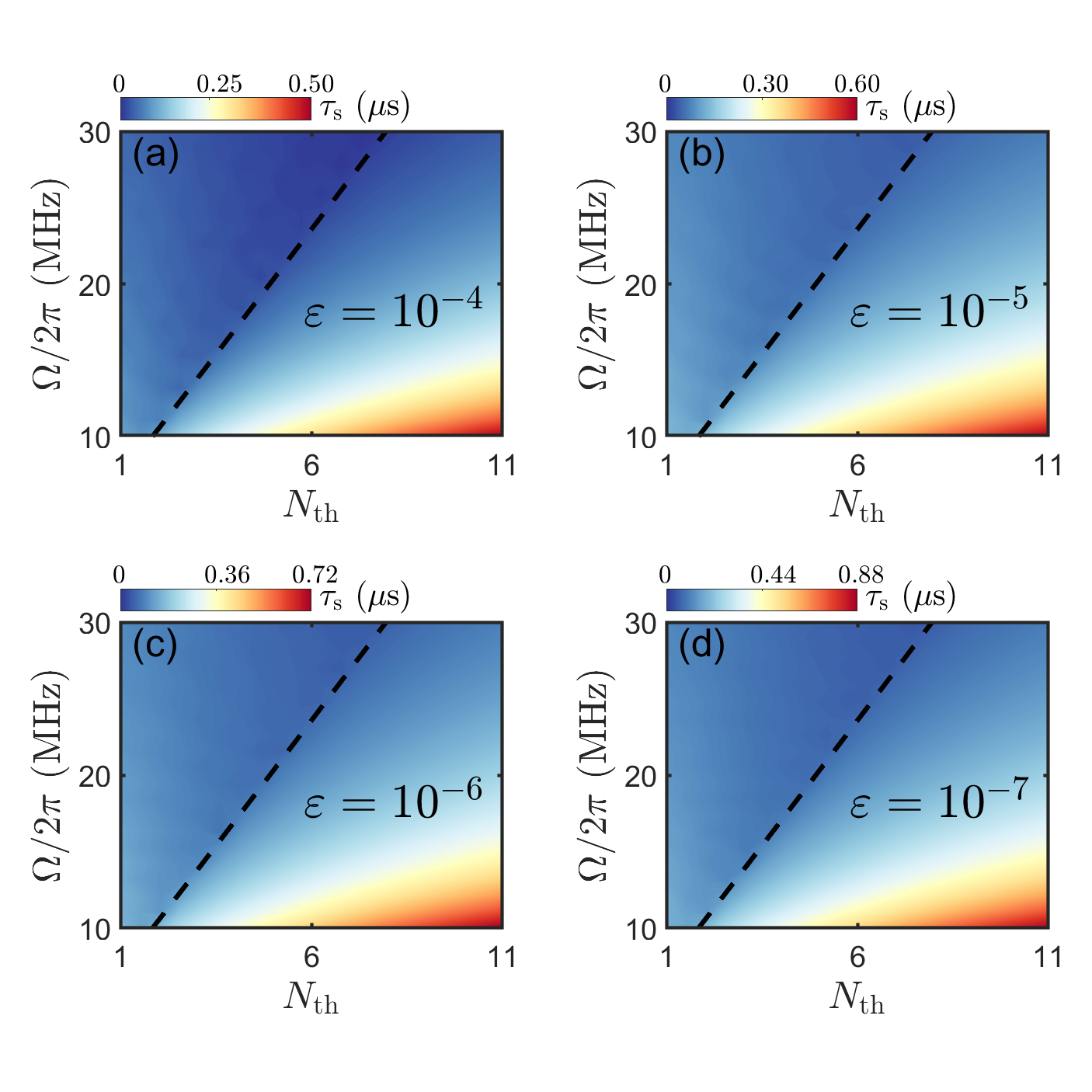}
        \caption{\label{fig7} Steady-state charging time $\tau_{s}$ in the
        $(N_{\mathrm{th}},\Omega)$ parameter plane for different convergence
        thresholds with $\delta/2\pi=0$: (a) $\varepsilon=10^{-4}$, (b)
        $\varepsilon=10^{-5}$, (c) $\varepsilon =10^{-6}$, and (d)
        $\varepsilon=10^{-7}$. The dashed line indicates the location of the EP.}
    \end{figure}
    Figure~\ref{fig7} shows a map of steady-state charging time $\tau_{s}$
    across the $(N_{\mathrm{th}},\Omega)$ parameter space for several
    convergence thresholds $\varepsilon$. Panels (a) to (d) display results for $\varepsilon
    =10^{-4}, 10^{-5}, 10^{-6}$, and $10^{-7}$. For all thresholds, rapid
    charging occurs at intermediate thermal occupations and strong driving
    strengths, while charging times increase with weak driving or large $N_{\mathrm{th}}$.
    The dashed line marks the exceptional-point trajectory identified by
    spectral analysis. Although reducing $\varepsilon$ changes the scale of $\tau
    _{s}$, the location and shape of the fast-charging region remain consistent.
    This confirms that the optimal charging regime is a robust feature of the system,
    regardless of the chosen threshold. The shortest $\tau_{s}$ coincides with the
    largest slow-sector gap, indicating that global charging performance is set
    by the spectral properties of the slowest-relaxing modes.\\
    \indent
    \begin{figure}[h]
        \centering
        \includegraphics[width=1\linewidth]{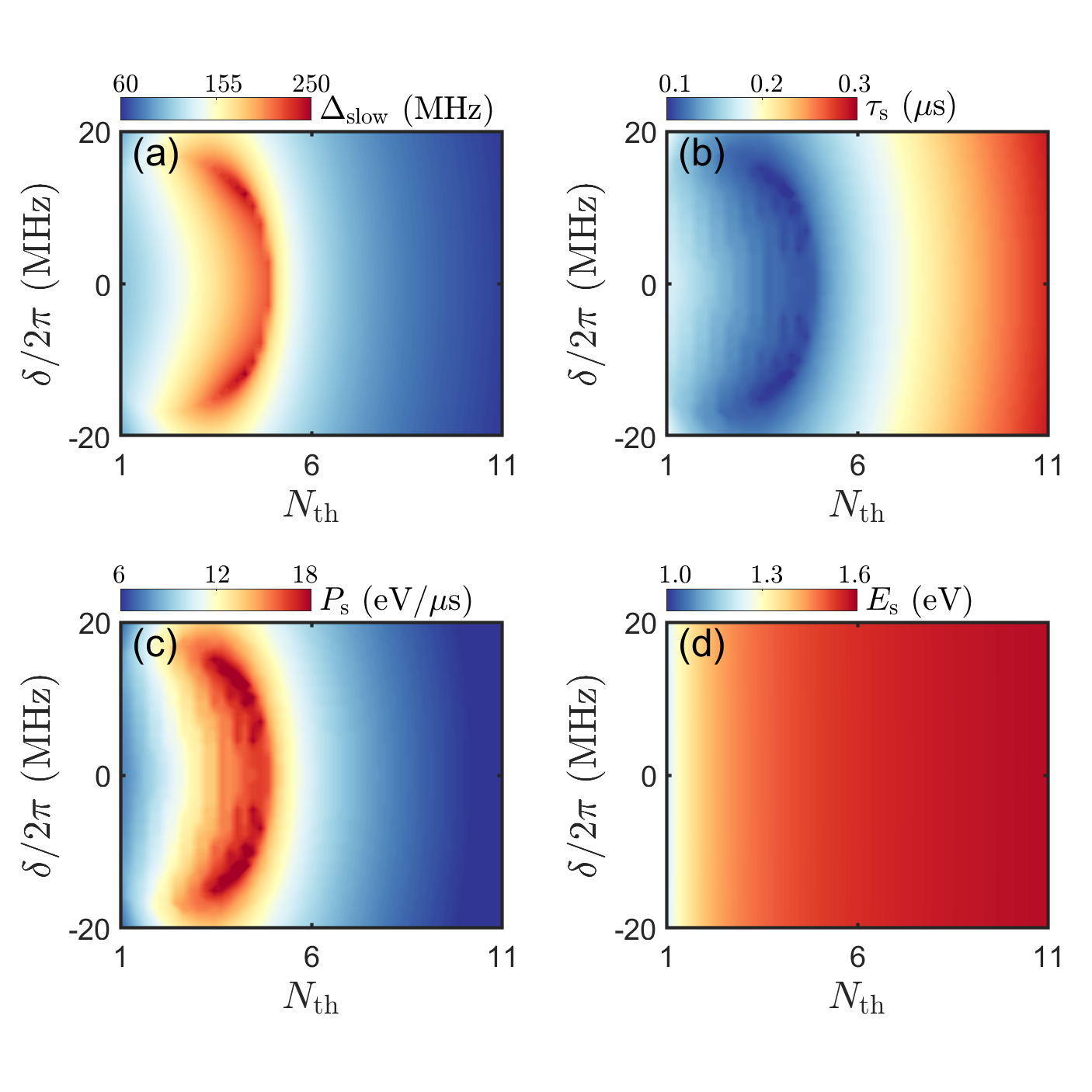}
        \caption{\label{fig8}Dependence on detuning in the $(N_{\mathrm{th}},\delta
        )$ parameter plane. (a) Slow-sector gap $\Delta_{\mathrm{slow}}$. (b) Saturation
        time $\tau_{s}$. (c) Steady-state charging power $P_{s}$. (d) Stored energy
        $E_{s}$. The parameters for the simulations
        are $\Omega/2\pi=20~\mathrm{MHz}$ and $\varepsilon = 10^{-8}$.}
    \end{figure}
    Figure~\ref{fig8} demonstrates that spectral and charging features persist
    under finite detuning $\delta$. In Fig.~\ref{fig8}(a), the slow-sector gap
    $\Delta_{\mathrm{slow}}$ forms a distinct, curved high-gap ridge near $\delta
    \approx 0$, reflecting the influence of EPs. This ridge determines the behavior
    in Fig.~\ref{fig8}(b), where the charging time $\tau_{s}$ is minimized, and in
    Fig.~\ref{fig8}(c), where the charging power $P_{s}$ is maximized along the
    same path. These optimal regions remain aligned as detuning shifts and bends
    the ridge, demonstrating robustness across the parameter space. In contrast,
    Fig.~\ref{fig8}(d) shows that the steady-state stored energy $E_{s}$ is
    largely unaffected by detuning, highlighting the dynamical nature of the observed
    enhancement. Finite detuning shifts the slow-sector reorganization away from
    exact resonance and smooths the eigenvalue structure, while preserving the slow-fast
    separation. Asymptotic relaxation remains governed by the leading nonzero
    Liouvillian eigenvalue, so the high-$\Delta_{\mathrm{slow}}$ ridge continues
    to mark the optimal charging region, even when displaced by $\delta$. These results
    show that while detuning changes the balance between coherent driving and dissipation,
    it does not eliminate the mechanism for optimal charging.\\
    \subsection{Experimental feasibility analysis}
    \label{Experimental_scope} The open QB charging model presented here is
    broadly applicable to other quantum platforms. This study focuses on its implementation
    in trapped-ion systems, using ${}^{40}\mathrm{Ca}^{+}$ ions as a
    representative example   \cite{RevModPhys.75.281,Barreiro2011,Myatt2000}. In this setup,
    coherent coupling between $|1\rangle$ and $|2\rangle$ is established with a
    resonant continuous-wave laser, where the Rabi frequency $\Omega$ is precisely
    controlled by adjusting the laser intensity. The detuning parameter $\delta$,
    representing the energy shift of $|2\rangle$ in the rotating frame, is
    finely tuned through the laser frequency
      \cite{PhysRevLett.56.2797,PhysRevLett.57.1699}. The engineered thermal reservoir
    between $|0\rangle$ and $|2\rangle$ is implemented using an incoherent optical
    field, with the effective mean occupation number $N_{\mathrm{th}}$ set by
    the relative rates of induced upward and downward transitions.\\
    \indent
    All key observables, including state populations, stored energies $E(t)$ and
    $E_{s}$, and the relaxation time $\tau_{s}$, are accessible using
    established techniques such as state-selective fluorescence, electron shelving,
    and time-resolved detection   \cite{PhysRevLett.56.2797,PhysRevLett.57.1699,PhysRevLett.111.023004}. The charging dynamics proceed on microsecond timescales,
    making decay from the metastable storage state negligible during experiments   \cite{PhysRevA.62.032503}.
    The parameters $(N_{\mathrm{th}}, \Omega, \delta)$ can be tuned independently   \cite{RevModPhys.75.281,PhysRevLett.77.4728,Myatt2000,Barreiro2011},
    enabling comprehensive experimental investigation of predicted effects, such
    as exceptional-point-enhanced spectral gaps and increased steady-state charging
    power.\\
    \indent
    Despite the high-fidelity control and readout offered by trapped-ion systems,
    several experimental challenges persist. These include the precise calibration
    of the engineered reservoir, suppression of undesired decoherence channels, and
    stabilization of laser parameters over the relevant timescales. Achieving
    robust incoherent coupling between $|0\rangle$ and $|2\rangle$ also requires
    effective isolation from environmental noise. Although these challenges can be
    addressed with current experimental methods, they require careful
    optimization to ensure reliable and reproducible results.\\
    \indent
    \section{Conclusion}
    \label{Conclusion} We theoretically demonstrated that steady-state charging
    in open quantum batteries (QBs) is fundamentally governed by the spectral
    properties of the Liouvillian superoperator. Using a minimal three-level trapped
    ${}^{40}\mathrm{Ca}^{+}$ ion as the QB model, we showed that the slowest
    Liouvillian modes dictate the long-time dynamics, with the spectral gap setting
    the relaxation timescale toward the steady state. Furthermore, we found that
    by tuning experimentally accessible parameters, the Liouvillian spectrum can
    be reshaped to enhance the dominant decay rate, thereby accelerating steady-state
    charging. This mechanism yields increased asymptotic charging power, arising
    from the spectral reorganization of slow-relaxation modes, without requiring
    many-body collectivity or steady coherence. Our results clarify that the
    structure of the slow-relaxation spectrum ultimately constrains the steady-state
    charging performance of open QBs. Therefore, Liouvillian spectral engineering
    offers a promising route to optimize dissipation-assisted energy storage by enabling
    the targeted design of decay modes that govern long-time dynamics.

    \begin{acknowledgments}
        This work was supported by the National Natural Science Foundation of China
        under Grants No. 12274470. H.Z. acknowledges partial support from the Postgraduate
        Scientific Research Innovation Project of Hunan Province under Grant No.
        2025ZZTS0117 and the Fundamental Research Funds for the Central
        Universities of Central South University under Grant No. 1053320240539. The
        simulation was conducted using computing resources at the High Performance
        Computing Center of Central South University.
    \end{acknowledgments}
    \appendix
    \section{Microscopic derivation of the interaction-picture Hamiltonian and
    Lindblad master equation}
    \label{supp1} The Hamiltonian for a three-level ion QB is
    \begin{equation}
        \hat{H}_{B}= \sum_{i=0}^{2}\omega_{i}|i\rangle\langle i|,
    \end{equation}
    where $|0\rangle$, $|1\rangle$, and $|2\rangle$ denote the electronic states.
    The photon reservoir is modeled as a collection of bosonic modes,
    \begin{equation}
        \hat{H}_{S}= \sum_{k}\nu_{k}\hat{a}_{k}^{\dagger}\hat{a}_{k}.
    \end{equation}
    The system--bath interaction that facilitates energy exchange between
    $|0\rangle$ and $|2\rangle$ is given by
    \begin{equation}
        \hat{H}_{SB}= \sum_{k}g_{k}\Bigl( |0\rangle\langle 2| + |2\rangle\langle
        0| \Bigr) \Bigl( \hat{a}_{k}+ \hat{a}_{k}^{\dagger}\Bigr).
    \end{equation}
    Additionally, a classical resonant control field couples states $|1\rangle$ and
    $|2\rangle$. Representing the field as $E(t) = E_{0}\cos(\omega_{L}t)$, the corresponding
    Hamiltonian is
    \begin{equation}
        \hat{H}_{\rm C}(t) = \mu_{12}E(t)\left[ |1\rangle\langle 2| + |
        2\rangle\langle 1| \right].
    \end{equation}
    The total Hamiltonian is thus
    \begin{equation}
        \hat{H}(t) = \hat{H}_{B}+ \hat{H}_{S}+ \hat{H}_{SB}+ \hat{H}_{\rm C}(t) .
    \end{equation}

    Transitioning to the interaction picture with respect to
    $\hat{H}_{0}= \hat{H}_{B}+ \hat{H}_{S}$ and applying the rotating-wave
    approximation, the system--bath coupling simplifies to
    \begin{equation}
        \hat{H}_{SB,I}(t) \simeq \sum_{k}g_{k}\left[ |0\rangle\langle 2|\, \hat{a}
        _{k}^{\dagger}e^{i(\omega_{20}-\nu_k)t}+ |2\rangle\langle 0|\, \hat{a}_{k}
        e^{-i(\omega_{20}-\nu_k)t}\right],
    \end{equation}
    where counter-rotating terms oscillating at $\omega_{20}+ \nu_{k}$ are
    neglected. The control field driving the $|1\rangle \leftrightarrow |2\rangle$
    transition is treated analogously, yielding the effective interaction-picture
    Hamiltonian
    \begin{equation}
        \hat{H}_{I}= \Omega \left( |1\rangle\langle 2| + |2\rangle\langle 1| \right
        ) + \delta |2\rangle\langle 2|,
    \end{equation}
    where $\Omega = \mu_{12}E_{0}/2$.

    The reduced dynamics of the ion are governed by the Born--Markov master equation
    \begin{multline}
        \frac{d\rho_{I}(t)}{dt}= -i[\hat{H}_{I}, \rho_{I}(t)] \\
        - \int_{0}^{\infty}d\tau\, \mathrm{Tr}_{R}\! \left[ \hat{H}_{SB,I}(t), \left
        [ \hat{H}_{SB,I}(t-\tau), \rho_{I}(t) \otimes \rho_{R}\right] \right], \label{eq:BM_master}
    \end{multline}
    where $\rho_{R}$ denotes the steady state of the engineered photon reservoir,
    characterized by mean occupation number $N_{\rm th}$. The relevant reservoir
    correlation functions are
    \begin{align}
        \langle \hat{a}_{k}^{\dagger}(t) \hat{a}_{k'}(t') \rangle & = \delta_{kk'}N_{\rm th}\, e^{i\nu_k(t-t')},      \\
        \langle \hat{a}_{k}(t) \hat{a}_{k'}^{\dagger}(t') \rangle & = \delta_{kk'}(N_{\rm th}+1)\, e^{-i\nu_k(t-t')}.
    \end{align}
    Collecting all contributions and returning to the Schr\"odinger picture yields
    the Lindblad master equation
    \begin{equation}
        \begin{aligned}
            \frac{d\rho(t)}{dt}={} & -i[\hat{H}_{I}, \rho(t)]+ \gamma_{21}\mathcal{D}[|1\rangle\langle 2|]\rho(t)+ \gamma_{10}\mathcal{D}[|0\rangle\langle 1|]\rho(t) \\                                                                             
                                   &  + \gamma_{20}(N_{\rm th}+1)\mathcal{D}[|0\rangle\langle 2|]\rho(t)+ \gamma_{20}N_{\rm th}\mathcal{D}[|2\rangle\langle 0|]\rho(t),
        \end{aligned}
        \label{eq:LME_final}
    \end{equation}
    which is the master equation presented in the main text.
    \section{Spectral Structure of the Five-Dimensional Liouvillian Block and
    Exceptional-Point Condition}
    \label{supp2} In the ordered operator basis
    $(\rho_{22},\rho_{12},\rho_{21},\rho_{11},\rho_{00})^{\mathsf{T}}$, the slow
    block is given explicitly by   \cite{PhysRevA.100.062131,PhysRevA.101.062112,PRXQuantum.2.040346}
    \begin{equation}
        \mathcal{L}_{5}=
        \begin{pmatrix}
            -2\Gamma                       & -i\Omega & i\Omega & 0        & N_{\mathrm{th}}\gamma_{20}  \\
            -i\Omega                        & -\Gamma & 0        & i\Omega  & 0                           \\
            i\Omega                       & 0       & -\Gamma  & -i\Omega & 0                           \\
            \gamma_{21}                    & i\Omega & -i\Omega & 0        & 0                           \\
            \gamma_{20}(N_{\mathrm{th}}+1) & 0       & 0        & 0        & -N_{\mathrm{th}}\gamma_{20}
        \end{pmatrix}, \label{eq:L5_matrix}
    \end{equation}
    where $\Gamma=\tfrac12\!\left[\gamma_{21}+\gamma_{20}(N_{\mathrm{th}}+1)\right
    ]$. It is convenient to introduce symmetric and antisymmetric coherence
    combinations $\sigma=\rho_{12}+\rho_{21}$ and $A=\rho_{12}-\rho_{21}$. From the
    equations of motion generated by Eq.~\eqref{eq:L5_matrix}, one finds that
    $\sigma$ evolves independently according to $\dot{\sigma}=-\Gamma\sigma$, so
    that $-\Gamma$ is an exact eigenvalue of $\mathcal{L}_{5}$. The remaining
    nontrivial slow dynamics is confined to the subspace spanned by
    $(\rho_{22},A ,\rho_{11}, \rho_{00})$.\\
    \indent
    Trace preservation implies that the physical density matrix satisfies
    $\rho_{22}(t)+\rho_{11}(t)+\rho_{00}(t)=1$ at all times, and consequently the
    Liouvillian $\mathcal{L}_{5}$ possesses an exact zero eigenvalue corresponding
    to the steady state $\rho_{\mathrm{ss}}$. To analyze relaxation dynamics, we
    consider deviations from the steady state,
    $\Delta\rho(t)=\rho(t )-\rho_{\mathrm{ss}}$, which obey the trace constraint
    $\mathrm{Tr}\,\Delta\rho=0$. In terms of components this constraint reads
    $\Delta \rho_{00}=-(\Delta\rho_{22}+\Delta\rho_{11})$.

    The slow relaxation dynamics is therefore governed by a closed three-dimensional
    linear system $\dot{\mathbf{x}}=M\mathbf{x}$ for the vector
    $\mathbf{x}=(\Delta \rho_{22},A,\Delta\rho_{11})^{\mathsf{T}}$. In the absence
    of detuning ($\delta /2\pi=0$), the effective generator reads
    \begin{equation}
        M=
        \begin{pmatrix}
            -(2\Gamma+N_{\mathrm{th}}\gamma_{20}) & -i\Omega & -N_{\mathrm{th}}\gamma_{20} \\
            -2i\Omega                             & -\Gamma  & 2i\Omega                    \\
            \gamma_{21}                           & i\Omega  & 0
        \end{pmatrix}. \label{eq:M_matrix}
    \end{equation}

    The three nonzero slow eigenvalues of $\mathcal{L}_{5}$ are therefore given by
    the roots of the cubic characteristic equation
    \begin{equation}
        \chi(\lambda)=\det(\lambda I-M)=\lambda^{3}+a\lambda^{2}+b\lambda+c=0. \label{eq:cubic_charpoly}
    \end{equation}
    After the standard shift $\lambda=x-a/3$, it reduces to the depressed cubic
    $x^{3}+p\,x+q=0$ with coefficients written compactly as $p=3P$ and $q=-2R$. For
    the matrix $M$ obtained from the trace-zero reduction of $\mathcal{L}_{5}$,
    one has
    \begin{equation}
        \begin{aligned}
            a & = \tfrac{1}{2}\!\left(5N_{\mathrm{th}}\gamma_{20}+3\gamma_{20}+3\gamma_{21}\right),                                                                            \\
            P & = \tfrac{4\Omega^{2}}{3}-\tfrac{(\gamma_{20}+\gamma_{21})^{2}}{12}-\tfrac{N_{\mathrm{th}}\gamma_{20}^{2}}{3}-\tfrac{13}{36}N_{\mathrm{th}}^{2}\gamma_{20}^{2}, \\
            R & = \tfrac{(3\gamma_{21}-4N_{\mathrm{th}}\gamma_{20})\Omega^{2}}{3}-\tfrac{N_{\mathrm{th}}\gamma_{20}(\gamma_{20}+\gamma_{21})^{2}}{24}                          \\
              & \quad -\tfrac{N_{\mathrm{th}}^{2}\gamma_{20}^{3}(36+35N_{\mathrm{th}})}{216}.
        \end{aligned}
        \label{eq:cubic_coeffs}
    \end{equation}
    The three solutions are given by the Cardano formula
    \begin{equation}
        x_{k}=\omega^{k}\sqrt[3]{R+\sqrt{\Lambda}}+\omega^{-k}\sqrt[3]{R-\sqrt{\Lambda}}
        , \qquad \omega=e^{2\pi i/3}, \label{eq:cardano_roots}
    \end{equation}
    and the corresponding Liouvillian eigenvalues are $\lambda_{k}=x_{k}-a/3$. The
    discriminant
    \begin{equation}
        \Lambda=\left(\frac{q}{2}\right)^{2}+\left(\frac{p}{3}\right)^{3}=R^{2}+P
        ^{3}\label{eq:discriminant}
    \end{equation}
    controls eigenvalue coalescence. In the present slow sector, the condition $\Lambda=0$ signals an
algebraic coalescence of two eigenvalues; together with the rank
condition $\dim\ker(M-\lambda_{\mathrm{EP}}I)=1$, this confirms a
defective degeneracy where the two eigenvectors also merge, realizing a
second-order Liouvillian EP.\\
\indent
To connect the EP condition with the local maximum of the slow-sector gap,
we examine the two dominant roots of the cubic equation near
\(\Lambda=0\). Along the tuning path \(N_{\mathrm{th}}\) at fixed \(\Omega\),
the coalescing eigenvalues take the standard square-root form
\begin{equation}
    \lambda_\pm
    =
    \lambda_{\mathrm{EP}}
    + c_{1}\left(N_{\mathrm{th}}-N_{\mathrm{th}}^{\mathrm{EP}}\right)
    \pm c_{2}\sqrt{N_{\mathrm{th}}-N_{\mathrm{th}}^{\mathrm{EP}}}
    +\cdots ,
    \label{eq:local_EP_expansion}
\end{equation}
where the coefficients \(c_{1}\) and \(c_{2}\) are obtained directly from the cubic
equation. For the parameter regime considered here, evaluating the roots gives
\(c_{1}<0\), and the remaining slow eigenvalue stays separated from this
coalescing pair. On the underdamped side
\((N_{\mathrm{th}}<N_{\mathrm{th}}^{\mathrm{EP}})\), the square-root term is
imaginary and therefore changes only the oscillation frequency, while the real
part moves closer to zero away from the EP. On the overdamped side
\((N_{\mathrm{th}}>N_{\mathrm{th}}^{\mathrm{EP}})\), the real splitting
produces a branch with a larger real part than \(\lambda_{\mathrm{EP}}\),
thereby reducing
\(\Delta_{\mathrm{slow}}=-\max_{\lambda_\beta\neq0}\mathrm{Re}[\lambda_\beta]\).
Thus, along this tuning direction, the EP gives a local maximum of
\(\Delta_{\mathrm{slow}}\).\\
\indent
    The behavior far away from the EP can be understood from the large-\(N_{\mathrm{th}}\)
limit. The
    relaxation toward the unique steady state is governed by the slow-mode
    spectral gap
    \begin{equation}
        \Delta_{\mathrm{slow}}\equiv -\max_{\beta\neq 0}\mathrm{Re}[\lambda_{\beta}
        (\mathcal{L}_{5})], \label{eq:gap_def_supp}
    \end{equation}
    where $\lambda_{\beta}(\mathcal{L}_{5})$ are the eigenvalues of
    $\mathcal{L}_{5}$ excluding the exact steady-state eigenvalue $\lambda=0$
    fixed by trace preservation. For $\delta=0$, the nontrivial slow spectrum is
    equivalently given by the three roots of $\det(\lambda I-M)=0$
    [Eq.~\eqref{eq:cubic_charpoly}] with $M$ in Eq.~\eqref{eq:M_matrix}. In the parameter regime of interest, the cubic equation has one real root and a complex-conjugate pair for $\Lambda>0$, corresponding to underdamped relaxation. At $\Lambda=0$, the dominant pair coalesces into a repeated eigenvalue; in the present non-Hermitian slow sector, the associated eigenvectors also coalesce, giving rise to a second-order EP. For $\Lambda<0$, the three roots are real, and the dominant pair splits into two distinct real decay rates, corresponding to overdamped relaxation.\\
    \indent
    The nonmonotonic dependence of $\Delta_{\mathrm{slow}}$ on $N_{\mathrm{th}}$
    follows from the competition between (i) the thermal exchange on the
    transition between $|0\rangle$ and $|2\rangle$ and (ii) the coherent
    transfer mediated by $\Omega$ between $|1\rangle$ and $|2\rangle$. In the reduced
    generator $M$ [Eq.~\eqref{eq:M_matrix}], the thermal occupation enters in
    two distinct ways: through the coherence damping scale $\Gamma$ and through the
    population-coupling matrix element $-N_{\mathrm{th}}\gamma_{20}$ that couples
    $\Delta\rho_{22}$ and $\Delta\rho_{11}$. Increasing $N_{\mathrm{th}}$
    therefore simultaneously increases the damping in the
    $(\rho_{12},\rho_{21})$ sector and the (non-Hermitian) coupling between
    population fluctuations. As a consequence, the dominant slow pair is pushed
    away from the imaginary axis, and the slowest decay rate $-\mathrm{Re}[\lambda
    _{\mathrm{slow}}]$ increases until the critical ``impedance-matching'' condition
    $\Lambda=0$ is met, where the underdamped pair collapses into a single
    defective mode. Precisely at EP, the slow sector loses one independent decay
    direction (Jordan-block formation), and the slowest decay rate is maximized,
    hence $\Delta_{\mathrm{slow}}$ reaches its peak.\\
    \indent
    For $N_{\mathrm{th}}$ beyond EP, thermal exchange on the $|0\rangle$--$|2\rangle$
    transition becomes parametrically large, and the slowest relaxation mode is no
    longer governed by the critical-damping condition at the EP. In the regime
    $N_{\mathrm{th}}\gamma_{20}\gg\{\Omega,\gamma_{21}\}$, the variables $(\Delta
    \rho_{22},A)$ constitute a fast sector and can be adiabatically eliminated,
    yielding an effective equation for the slow storage population
    $\Delta \rho_{11}$. Setting $\dot{\Delta\rho}_{22}\simeq0$ and $\dot{A}\simeq
    0$ in Eq.~\eqref{eq:M_matrix} and solving for $(\Delta\rho_{22},A)$ in terms
    of $\Delta\rho_{11}$ gives
    \begin{equation}
        \begin{aligned}
            \Delta\rho_{22} & = \frac{2\Omega^{2}-\Gamma N_{\mathrm{th}}\gamma_{20}}{2\Gamma^{2}+\Gamma N_{\mathrm{th}}\gamma_{20}+2\Omega^{2}}\,\Delta\rho_{11},        \\[4pt]
            A               & = \frac{4\mathrm{i}\Omega(\Gamma+N_{\mathrm{th}}\gamma_{20})}{2\Gamma^{2}+\Gamma N_{\mathrm{th}}\gamma_{20}+2\Omega^{2}}\,\Delta\rho_{11}.
        \end{aligned}
        \label{eq:adiabatic_solution}
    \end{equation}
    Expanding for large $N_{\mathrm{th}}\gamma_{20}$ yields
    \begin{equation}
        \begin{aligned}
            \Delta\rho_{22} & \simeq \left(-\frac{1}{2}+\frac{\gamma_{20}+\gamma_{21}}{4N_{\mathrm{th}}\gamma_{20}}\right)\Delta\rho_{11},                              \\[4pt]
            A               & \simeq \frac{6\mathrm{i}\Omega}{N_{\mathrm{th}}\gamma_{20}}\,\Delta\rho_{11}+\mathcal{O}\!\left((N_{\mathrm{th}}\gamma_{20})^{-2}\right).
        \end{aligned}
        \label{eq:adiabatic_solution_asymptotic}
    \end{equation}
    Substituting Eq.~\eqref{eq:adiabatic_solution} into $\dot{\Delta\rho}_{11}=\gamma
    _{21}\Delta\rho_{22}+\mathrm{i}\Omega A$ produces an effective single-mode relaxation
    $\dot{\Delta\rho}_{11}\simeq-\kappa_{\mathrm{eff}}\,\Delta\rho_{11}$, with
    \begin{equation}
        \kappa_{\mathrm{eff}}=\frac{\Gamma N_{\mathrm{th}}\gamma_{20}\gamma_{21}+4\Omega^{2}(\Gamma+N_{\mathrm{th}}\gamma_{20})-2\Omega^{2}\gamma_{21}}{2\Gamma^{2}+\Gamma
        N_{\mathrm{th}}\gamma_{20}+2\Omega^{2}}. \label{eq:kappa_eff_general}
    \end{equation}
    Using $\Gamma=\frac{1}{2}[\gamma_{21}+\gamma_{20}(N_{\mathrm{th}}+1)]$, one obtains
    \begin{equation}
        \kappa_{\mathrm{eff}}=\frac{\gamma_{21}}{2}+\frac{24\Omega^{2}-\gamma_{21}(\gamma_{20}+\gamma_{21})}{4N_{\mathrm{th}}\gamma_{20}}
        +\mathcal{O}\!\left((N_{\mathrm{th}}\gamma_{20})^{-2}\right), \label{eq:kappa_eff_asymptotic}
    \end{equation}
    so that the slowest nonzero eigenvalue behaves as $\lambda_{\mathrm{slow}}\simeq
    -\kappa_{\mathrm{eff}}$ and the gap satisfies
    \begin{equation}
        \Delta_{\mathrm{slow}}\simeq\kappa_{\mathrm{eff}}\;\longrightarrow\;\frac{\gamma_{21}}{2}
        \qquad(N_{\mathrm{th}}\to\infty). \label{eq:gap_largeNth}
    \end{equation}\\
    \indent
    Thus, beyond EP the gap decreases from its maximal value at the EP but remains
    finite in the large-$N_{\mathrm{th}}$ limit. Physically, strong thermal exchange
    rapidly mixes the $\{|0\rangle,|2\rangle\}$ manifold, while the residual
    slow relaxation is ultimately limited by the irreversible decay channel
    associated with $\gamma_{21}$. The post-EP slowdown therefore reflects a spectral
    reorganization of the slow sector rather than a vanishing $\Omega^{2}/(N_{\mathrm{th}}
    \gamma_{20})$ rate.

\end{document}